\documentclass[prl,amsmath,amssymb,superscriptaddress,twocolumn,floatfix]{revtex4}
\usepackage{graphicx}

\renewcommand{\Psi}{\varPsi}

\begin{document}

\title{Precision cosmological measurements:  independent evidence for dark energy}

\author{Greg~Bothun} \email{bigmoo@gmail.com}
\affiliation{Department of Physics, University of Oregon,
Eugene, OR 97403}

\author{Stephen~D.~H.~Hsu} \email{hsu@uoregon.edu}
\affiliation{Department of Physics, University of Oregon,
Eugene, OR 97403}
\affiliation{Institute of Theoretical Science, University of Oregon,
Eugene, OR 97403-5203}

\author{Brian~Murray}
\email{bmurray1@uoregon.edu}
\affiliation{Department of Physics, University of Oregon,
Eugene, OR 97403}
\affiliation{Institute of Theoretical Science, University of Oregon,
Eugene, OR 97403-5203}

\begin{abstract}
Using recent precision measurements of cosmological paramters,
we re-examine whether these observations alone, independent of type Ia supernova
surveys, are sufficient to imply the existence of dark energy. We find
that best measurements of the age of the universe $t_0$, the Hubble
parameter $H_0$ and the matter fraction $\Omega_m$ strongly favor an
equation of state defined by ($w < -1/3$).  This result is consistent with the
existence of a repulsive, acceleration-causing component
of energy if the universe is nearly flat.
\end{abstract}



\maketitle


The current era in cosmology seems to be the first in which local astrophysical measurements
are consistent with the generally accepted large scale cosmology. 
To provide some historical context, consider the period from 1980 to roughly 1995.  Inflation offered us a large scale model for cosmology, requiring $\Omega_{\rm total}  = 1$,  which could not
find verification in measurements on smaller scales. Attempts to dynamically determine $\Omega_{\rm total}$ (e.g., \cite{A1986,B1992}) consistently returned results of $\Omega_{\rm total} \sim 0.25 \pm 0.10$.  This led to the notion \cite{SW1995} that, under the $\Omega_{\rm total}  =1$ prior, there must be a bias between the distribution of light (e.g. galaxies) and mass (e.g. the dark matter component).  Not only did the Univese have to be dark matter dominated, the distribution
of that dark matter had to be signficantly different than the the
distribution of light.  At the time, this was the only way to reconcile the small
scale measurements with the large scale (inflation) requirement.  
 
In this note we reinvestigate whether recent determinations of cosmological 
parameters are sufficient, by themselves, to imply the existence of dark energy -- 
specifically, a component of energy with equation of state $w \equiv p / \rho < -1/3$.
In the mid-90's several authors \cite{early,B1998} analyzed aggregate
data based on globular cluster ages, clustering of galaxies, big bang nucleosynthesis, and
the Hubble constant and concluded that something like a cosmological constant might be necessary
to produce a flat Universe.  However, the conclusions were not definitive at the time
due to the large uncertainty in the observational parameters. Our purpose is to 
update these earlier investigations, accounting for improvements in precision. 
We will argue that observations of key parameters such as the age of the universe $t_0$,
the Hubble parameter $H_0$ and the matter fraction $\Omega_m$ 
have become definitive in support of dark energy. One might question the need
for this analysis in the post-WMAP era, but it is important to understand whether
increasingly precise measurements are consistent with the concordance cosmology
obtained from best fits of WMAP data. Indeed,
given the dramatic nature and consequences of dark energy, it is
important to understand the observational evidence for it as broadly
and robustly as possible.  

Despite the impressive results of the type
Ia supernova collaborations \cite{SN}, it is still possible that dust
\cite{dust}, evolution effects \cite{evo} or exotic particle physics
\cite{axions} might alter the interpretation of the extracted
redshift-distance relation.  For example, the axion models in
\cite{axions} account for the dimness of distant supernovae by
conversion of photons into axions in background galactic magnetic
fields, rather than through accelerated expansion. Exotic particle physics models 
which are less well motivated than axions, but perhaps no more counterintuitive than 
the existence of dark energy itself, might in principle explain the supernova data 
without requiring acceleration.  However, the demonstration that a dominant component
of energy with  
$w \equiv p / \rho < -1/3$ is strongly favored by the observed values of cosmological
parameters provides a direct and robust argument for acceleration.


We seek evidence for a component
which has equation of state $w \equiv p / \rho < -1/3$. Recall the
Einstein equation
\begin{equation}
\label{EE}
\frac{\ddot{R}}{R} = -\frac{4 \pi G}{3}\sum_i (\rho_i + 3 p_i)~.
\end{equation}
The sign of the acceleration $\ddot{R}$ is determined by the sign of
$\sum_i (\rho_i + 3 p_i )$, where the sum runs over all contributions
to the energy momentum tensor. Strictly speaking, $w < -1/3$ is the
threshold for a component to cause acceleration when it is the
only form of energy. If other forms of energy are non-negligible the
overall sign of the right hand side of (\ref{EE}) might still be
negative (i.e., the universe is decelerating, albeit more slowly than
otherwise) even in the presence of energy with $w <
-1/3$. Asymptotically, though, the component with the smallest
positive or most negative value of $w$ will eventually dominate all
others. We recall that a cosmological constant has $w = -1$, while a
dynamical scalar model with non-zero vacuum energy typically has $-1 <
w < 0$. Values of $w$ less than $-1$ violate the null energy
condition, and are generally associated with instabilities \cite{nec}.

Analysis of the 3 year WMAP data \cite{WMAP} favors a negative
pressure equation of state for models with constant $w$ when
constraints on the matter energy density are included (i.e., from the
Sloan Digital Sky Survey or the 2dF Galaxy Redshift Survey). In this note 
we conduct a simpler analysis in
which the priors are transparent and easy to state.

We find that best measurements of the age of the universe $t_0$, the
Hubble parameter $H_0$ and the matter fraction $\Omega_m$ are
sufficient to require the existence, during some cosmologically
significant epoch, of a repulsive, acceleration-causing ($w < -1/3$)
component of energy, assuming the universe is nearly flat. A relation between these quantities is obtained using Einstein's equation for a Friedmann-Robertson-Walker universe. The analysis itself is not necessarily new, but it can now be applied for the very first time with stringent constraints due to recent precision measurements of the relevant cosmological parameters.

The age of the universe is given by
\begin{equation}
t_0 = \int_0^{R(t_0)} \frac{dR}{\dot{R}} 
\end{equation}
which yields
\begin{equation}
\label{int}
t_0 H_0 = \int_0^1 \frac{dx}{ ( \Omega_m x^{-1} + \Omega_{de}
x^{-1-3w} )^{1/2} } ~,
\end{equation}
where we have taken $w$ constant in time and neglected the radiation
component as it is numerically small. We also assume flatness, which
implies $\Omega_{de} = 1 - \Omega_m$, and allows us to define the
integral as $I(\Omega_m , w)$. The quantities $t_0$, $H_0$ and
$\Omega_m$ then determine $w$.

In the more general case, where the dark energy component has time
varying equation of state $w(t)$, the second term in the denominator
of the integral in (\ref{int}) (the dark energy term) is more
complicated, having the form
\begin{equation}
\label{de}
\Omega_{de} \exp \left[ \int_x^1 \frac{dx'}{x'} (1 + 3 w(x')) \right]~.  
\end{equation}
If $(1 + 3 w(x')) > 0$ for all $x< x' < 1$, the dark energy term
(\ref{de}) is always decreasing with increasing $x$, and the
denominator in (\ref{int}) is larger for all $x$ than it would be in
the special case $w = -1/3$, where (\ref{de}) is constant. Therefore,
if the dark energy {\it never} exhibits a repulsive equation of state,
so $w(t) > -1/3$ at all times, the integral is bounded above:
\begin{equation}
I (\Omega_m, w > -1/3 ) ~<~ I (\Omega_m, -1/3)~.
\end{equation}
Similarly, we deduce
\begin{equation}
\label{inq}
I (\Omega_m, w > w^* ) ~<~ I (\Omega_m, w^*)~.
\end{equation}
In other words, in the most general case, unless the dark energy
behaved repulsively during some earlier epoch, the integral $I$, and
hence the product $t_0 H_0$, is bounded above by $I (\Omega_m,
-1/3)$. Using measured values of $t_0$, $H_0$ and $\Omega_m$, it is
therefore possible to deduce that a repulsive epoch must have
occurred. (Note an epoch with repulsive energy does not necessarily
imply overall acceleration, as discussed.)


We now review the best measurements of $t_0$, $H_0$ and
$\Omega_m$. Systematically combining the results of distinct
measurements using different techniques, each with different
statistical and systematic errors, is challenging. However, our
discussion at least allows a reasonable guess at current global best
values and uncertainties for these quantities. 
Examples of more sophisticated Bayesian analysis are given in \cite{Bayes}.

{\bf $t_0$:} Our approach is made possible by relatively recent
measurements of $t_0$ with unprecedented accuracy. In the past,
estimates of $t_0$ have been made by either using model-dependent
estimates for the ages of globular clusters or through nuclear
cosmochronometry. The former method has traditionally suffered from
the unknown role of convection and its effects on the lifetimes of low
mass/low metallicity stars.  Krauss and Chaboyer \cite{KC} performed a
thorough Monte Carlo analysis that includes these uncertainties, to
arrive at a firm lower limit of 11.2 Gyrs for $t_0$.  However, $t_0$
as large as 15 Gyrs is still allowable.  Using Thorium
cosmochronometry, Sneden and Cowan \cite{SC} also find a lower limit
of 11 Gyrs for $t_0$ but acknowledge that lower limit could range
upwards by another 3-4 Gyrs.  For the reasons cited, we do not use these
methods or observations in construction our argument for the most probable
value of $t_0$.

Improvements in the precision of
measuring $t_0$ have utilized the white dwarf cooling curve and Hubble
Space Telescope measurements of the halo globular cluster M4.
Measurements by Hansen et al. (2002) \cite{Hanson2002} report a value
of $12.7 \pm 0.7$ Gyr.  Hansen et al. (2004) \cite{Hanson2004} update
this age to $12.1 \pm 0.9$ Gyr.  The major source of systematic error
in this analysis involves estimating the lag time between the age
of the Universe and the formation of globular clusters.
Numerical simulations of the Milky
Way and its globular cluster system by Kravtsov and Gnedin (2005)
\cite{Gnedin} indicate that the peak formation of Globular Clusters
occurs at $z =$ 3-5.  Using a mean formation redshift of $z=4$ implies
that Globular Clusters formed at 1.2 Gyr after the onset of the Big
Bang.  This then leads to a lower limit of $t_0$ = 12.4 Gyr and a mean
value of $t_0 = 13.3^{+1.1}_{-.9}$ Gyr.

\bigskip

{\bf $H_0$:} For decades, measurements of $H_0$ were plagued by noise
and biased samples.  Today, however, there is good reason to believe
that we have a relatively precise measure for this parameter as well.
The Hubble Space Telescope Key Project for determining the Cepheid
Zero Point and subsequent distance determinations to nearby galaxies
using the Cepheid Period-luminosity relationship have returned a value
of $72 \pm 3$ km/s/Mpc \cite{HST}.  The major source of systematic
uncertainty in that measurement lies in the distance to the Large
Magellanic Cloud (LMC), to which the zeropoint of the Cepheid
Luminosity scale is anchored.  Freedman and Madore \cite{FM} quote a
total systematic error of $\pm 7$ km/s/Mpc, but recent improved
distance estimates for the LMC (e.g., Benedict et al. (2002) \cite{Ben}
and Sebo et al. \cite{Seb}) have served to lower this systematic error
down to $\pm 4$ km/s/Mpc (see Ngeow and Kangur (2006) \cite{NK}).
Moreover, confidence in the precision of $H_0$, as anchored by the 
LMC distance, is reinforced by recent
measurements that are completely independent of the distance
to the LMC.  In the past, these kinds of measurements were also available
but they had sufficiently large random error that precluded them from
providing meaninful constraints on the value of $H_0$ as determined from
traditional distance scale ladder techniques.  The new observations are:

1) Using a sample of 38 X-ray clusters in combination with the
Sunyaev-Zeldovich effect, Bonamente et al. (2006) \cite{Bon} derive a
value of $H_0 = 77.6 \pm 5$ km/s/Mpc.  While there may be systematics
associated with the non-spherical shape of clusters, their sample may
is sufficiently large (and much larger than past samples)
that this problem is removed by averaging.

2) Wang et al. (2006) \cite{Wan} have examined a sample of 109 SN of
type Ia and have discovered important new corrections for metallicity
and absorption (by dust) in determining SN Ia peak luminosity.  This
recalibration leads to $H_0 = 72 \pm 6$ km/s/Mpc.  An independent
treatment of SN Ia has been compiled by Riess et al. (2005) \cite{Rie}
which yields a value of $H_0 = 73 \pm 4$ km/s/Mpc with possible
systematic error of $\pm 5$ km/s/Mpc.  

3) Koopmans et al. (2003)
\cite{Koo} perform a detailed analysis of a gravitational lens system
(from which a direct determination of the distance can be determined
using a model mass distribution of the lens) to find $H_0 = 75 \pm
6.5$ km/s/Mpc.

Averaging these 5 different results together formally leads to $74 \pm
2.5$ km/s/Mpc (error in the mean). Direct averaging is crude, but
gives a characterization of the uncertainty. Averaging over systematic
errors as well, we assume $H_0 = 74 \pm 5$ km/s/Mpc in further
analysis.  In contrast, one could use only method 1 and 3 above (as
they completely circumvent the LMC distance problem) to obtain 76 $\pm$
6 as the relevant range.

\bigskip

{\bf $\Omega_m$:} In contrast, $\Omega_m$ remains the most weakly
constrained cosmological observable.  There are two reliable methods
of measurement: dynamical determinations based on infall to clusters
of galaxies and/or the nature of large scale structure (e.g., Bothun et
al. \cite{Bot}) or by fitting the Hubble diagram to distant
objects.  In the first case, an unbiased and fairly large sample is
needed for precision; in the second case, accurate distance
measurements of intermediate redshift galaxies are required, and such
measurements are ultimately based on the supernova luminosity scale.
In principle, $\Omega_m$ is highly constrained by the multi-parameter
maximum likelihood fit to the WMAP data; but this is an indirect
determination  of $\Omega_m$ (as well as $t_0$)  In the spirit of this analysis,
we seek to use values of $\Omega_m$ that have been directly determined.

Note, though, that $\Omega_m$ is now
usually determined by assuming a flat Universe as a prior constraint. 
For instance, a recent
accurate determination of $\Omega_m$ results from analysis of the
power spectrum of galaxy clustering.  Assuming a flat Universe,
Sanchez et al. (2006) \cite{San} find $\Omega_m = 0.237 \pm 0.02$.  In
addition, Mohayee and Tully (2005) \cite{MT} revisit the peculiar
velocities of galaxies in the Local Supercluster to derive $\Omega_m =
0.22 \pm 0.02$.  Schindler (2002) \cite{Sch} summarizes all techniques
to determine $\Omega_m$ (including the more unreliable approaches such
as the X-ray cluster luminosity function, weak gravitational lensing,
or galaxy cluster evolution).  That summary yields a modal value of
$\Omega_m =0.3$ (which is likely a realistic upper limit given the
WMAP model) but also shows that most large scale structure studies
yield values of $\Omega_m$ in the range 0.20 - 0.25 (which is
consistent with the work done in the 1980s).  Averaging together
the Sanchez et al. and Mohayee and Tully studies produces a well
constrained value of $\Omega_m$ = 0.23 $\pm$ 0.02.
For discussion below we take a conservatively large range for
$\Omega_m$, assuming $0.15 - 0.25$ to be a one standard
deviation range about the central value.

\bigskip

\begin{figure}[ht]
\includegraphics[width=8cm]{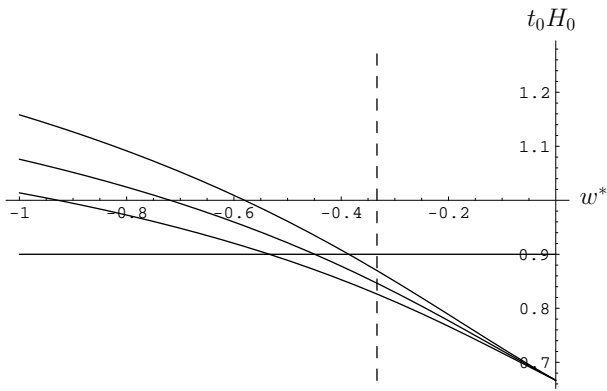}
\caption{Curves in the $w - t_0 H_0$ plane, each of which is an {\it upper} bound on
$t_0 H_0$, for $\Omega_m = .25, .20, .15$. The allowed region is between the top and
bottom curves, and above the horizontal line $t_0 H_0 = 0.9$. This requires $w$ less than $-1/3$.}
\label{figure1a}
\end{figure}

{\bf Results:} In Fig. \ref{figure1a}, we plot $I( \Omega_m, w)$ for
$\Omega_m = .15, .20$ and $.25$. $\Omega_m = .15$ corresponds to the
curve with the largest values of $t_0 H_0$. Taking $t_0 = 12.4$ Gyr
and $H_0 = 69$ km/s/Mpc, which are each one standard deviation below
the favored (central) values in our assumed error model, we obtain
$t_0 H_0 = .9$, which corresponds to the grey horizontal line in the
figure. The implications can be read directly from the figure. If $w$
was always greater than $-1/3$, then some or all of our parameters
must be well below their central values.

From Fig. \ref{figure1a}, we see that taking $t_0$, $H_0$ and
$\Omega_m$ to each be one standard deviation below their central value
(so, $t_0 = 12.4$ Gyr, $H_0 = 69$ km/s/Mpc and $\Omega_m = .15$), an
epoch with $w < -.4$ or so is required, which is just negative enough
to imply acceleration ($\ddot{R} > 0$). Taking $t_0 = 12.4$ Gyr and
$\Omega_m = .15 \,$, one would have to, e.g., push $H_0$ below 67
km/s/Mpc to have $w > -1/3$, and below 50 km/s/Mpc to have $w > 0$ (no
negative pressure).

We compute the likelihood of no epoch with $w < w^*$ (for given $w^*$) as follows. First, we assume uncorrelated Gaussian errors in all three
parameters: $t_0 = 13.3 \pm 1$ Gyr, $H_0 = 74 \pm 5$ km/s/Mpc and
$\Omega_m = .2 \pm .05$ (all one standard deviation).  That is, we assume that the probability distribution for the actual value each of parameter  is normal, with maximum at the central value and standard deviation given by the error estimate. We then compute, for a particular value of $w^*$, the total probability that the parameters take on values for which inequality (\ref{inq}) is satisfied. In practice, this was done using Monte Carlo.

\begin{figure}[ht]
\includegraphics[width=8cm]{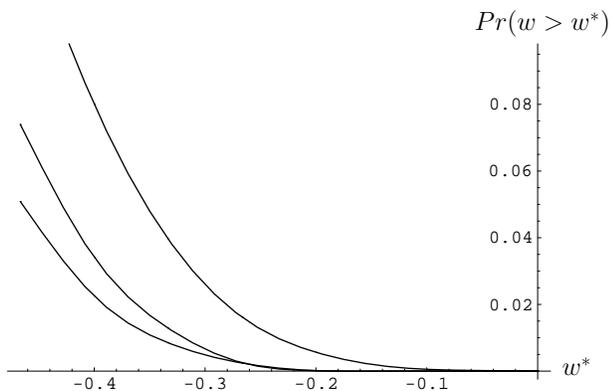}
\caption{Probability that $w$ was always greater than $w^*$ for a
range of $w^*$ and various cuts on $t_0$ and $\Omega_m$. See text for
details.}
\label{figure2a}
\end{figure}
\begin{figure}[ht]
\includegraphics[width=8cm]{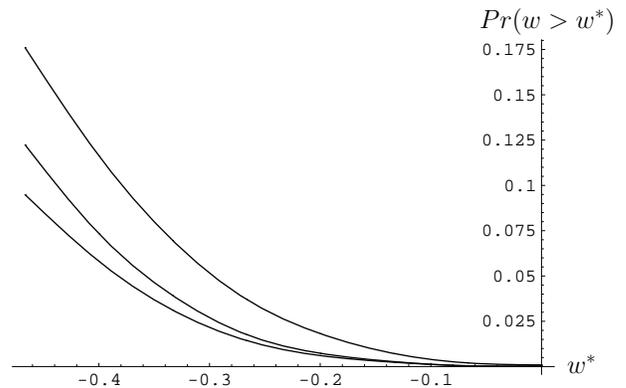}
\caption{Same as Fig. \ref{figure2a} except with larger Hubble
  uncertainty: $H_0 = 74 \pm 7$ km/s/Mpc.}
\label{figure3a}
\end{figure}

The results are displayed in
Fig. \ref{figure2a} (top curve). Using this error model the
probability of no epoch with $w < -1/3$ is less than 4 percent.  This
is an {\it overestimate} of the likelihood, since the model allows values
of, e.g., $t_0$ which are much too low: $t_0 = 12.4$ Gyr is more
plausibly interpreted a strict minimum than minus one standard
deviation from the central value.  Modifying the error model so that
values of $t_0 < 12.4$ Gyr are not allowed reduces the likelihood of
no epoch with $w < -1/3$ to about 1.3 percent. This is represented by
the middle curve in Fig. \ref{figure2a}.  Adding a similar constraint
that $\Omega_m > .15$ leads to the lowest curve in the figure, and a
likelihood of no epoch with $w<-1/3$ of about 0.8
percent. Fig. \ref{figure3a} is identical to Fig. \ref{figure2a}
except that we have increased the one standard deviation error for
$H_0$ to $\pm$ 7 km/s/Mpc; the existence of dark energy is still
strongly favored.

We conclude that, unless systematic errors are significantly larger than  currently recognized, best measurements of the age of the universe
$t_0$, the Hubble parameter $H_0$ and the matter fraction $\Omega_m$
strongly favor the existence of a repulsive dominant energy component, also
known as dark energy. These observations are independent of type Ia
supernova surveys: specifically, they are not sensitive to
uncertainties \cite{dust,evo,axions} which affect the direct
measurement of the distance-redshift relation at large $z$.

\bigskip

\begin{acknowledgments}
\section{acknowledgements}
The authors thank P. Corasaniti for clarification of his work in \cite{dust}. S.~H. and B.~M. are supported by the Department of Energy under
DE-FG02-96ER40969.
\end{acknowledgments}




\end{document}